\documentclass[12pt]{article}
\usepackage{epsfig}

\def\parno {\par\noindent}
\begin{document}
%\centerline{\large\bf photondrag\_30\_10\_02.tex}
%%
%
\begin{flushright}
{\bf \small IFUM 724/FT}
\end{flushright}

\vskip 1 truecm
%\rightline{Milan,  October 30, 2002 }
%\rightline{revised   October 30, 2002}
\vskip 1 truecm
%\magnification=\magstep1
\tolerance=100000
\centerline {\bf Virtual Photon Contribution}
\centerline {\bf to Frictional Drag in double-layer Devices}
\vskip 1 truecm
\centerline {A.Donarini${}^1$, R.Ferrari${}^2$, A.P.Jauho${}^1$,
L.Molinari${}^2$}
\centerline {1) Mikroelektronik Centret, Bygning 345 {\o}st, DK-2800 Lyngby}
\centerline {2) Dipartimento di Fisica and I.N.F.N., Via Celoria 16,
I-20133 Milano}

\vskip 0.5truecm
{\bf {Abstract.}} The first order contribution to
frictional drag in  bi-layered fermion gas is examined.
We discuss the relevance of single photon exchange
in the evaluation of transresistance,
which is usually explained by second order effects such as Coulomb and
phonon drag. Since the effective e.m. interaction is unscreened,
in the d.c. limit we obtain a finite (and large)
contribution to transconductivity.
\vskip 1truecm

PACS codes: 73.43.-f, 73.21.-b

\vfill\eject
\section{Introduction}
\label{sec:intro}
The double layer configuration of the fermion gas, especially under the
influence of a strong magnetic field, allows one to study
unusual transport phenomena which are interesting for the theory and are
nowadays accessible to experiments. In 1976, Pogrebinskii and Price advanced
the idea that a current driven in an electron
gas should manifest as a potential difference in a separated electron system
because of Coulomb scattering with a preferred direction of exhanged momentum.
This Coulomb drag was described in terms of coupled transport
equations. Later on the effect, which was originally analyzed for bulk
electrons, was experimentally observed in double well heterostructures at low
temperature,  allowing for quantitative measurements \cite{Gra}.\parno
The measured quantity is transresistance, the ratio of potential
difference due
to dragged charges and driving current. It was the main subject of several
theoretical works based on transport equations or Kubo formula for
conductivity \cite{Kam,Flens}. Transresistance resulting from Coulomb drag depends
on temperature as $T^2$ and decreases as $d^{-4}$ in the distance between the
wells. The theory was refined by including phonon exchange \cite{Tso,Bons},
which gave
the main explanation of the observed deviations from the $T^2$ behaviour.
A "current drag" effect was also proposed \cite{Mah2}, and originates from
the Van der Waals attraction between relative current flows.\parno
With strong magnetic field, the bilayer geometry allows the observation
of new FQHE phenomena and is an active research area \cite{Rojo}.\par
In the Kubo formalism transconductivity is {given by }  the retarded
correlator of currents in different layers. As a consequence of charge
conservation, it is a second order quantity in the electrostatic and
phonon interlayer interaction. In this letter we discuss the effect of photon
exchange, in the absence of external magnetic field, by considering  the
coupling of the photon field to currents, in the
Coulomb gauge.  The inclusion of photon exchange seems to us quite natural,
given that photon drag arises as a first order effect. Though e.m. corrections
are usually negligible, it was not obvious to us how they compared with
second order effects.\parno
The e.m. coupling in electron gas is the source of  small though
interesting effects. The  effective e.m. interaction was evaluated in
R.P.A. by Holstein et al. \cite{Hol} and displays the main feature of being
unscreened at zero frequency. Reizer \cite{Rei} investigated its influence
on the Fermi surface, the low temperature specific heat, and homogeneous
transport. Gauge-invariant response functions were studied by Kim et al.
\cite{Kim}, and confirmed the Fermi liquid behaviour. \parno
In this letter the photon polarization
is evaluated in R.P.A. in the interlayer e.m. interaction, in the
limit  of thin layers. Single layer properties and disorder are
accounted for in the diffusive regime. The results are discussed in the
concluding section.
\vskip 1truecm
We model the system as two infinite parallel layers of electron
gas, confined in narrow potential wells centered in $z=0$ and $z=d$, with
negligible tunnelling because of low temperature and layer separation.
The fermions in the two layers can be described by independent field
operators.
\parno
The Hamiltonian is
\begin{eqnarray}
H=H_1+H_2 +H_{ph} +U_{Cou}+U_{em},
\label{(0.1)}
\end{eqnarray}
where  $H_\ell $ is the
kinetic  energy of the electrons in layer $\ell=1,2$,
$H_{ph}$ is the energy of free e.m. field, $U_{Cou}$ is the
Coulomb interaction. The e.m. minimal coupling with the vector
potential in the Coulomb gauge is
\begin{eqnarray}
H_{em}= {1\over c}\int d^3x {\underline j} (\underline x) \cdot
 {\underline A} (\underline x)
- {e\over {2mc^2}}\int d^3x \rho (\underline x){\underline A}^2
(\underline x)
\label{(1)} \end{eqnarray}
where $\rho $ is the electron charge density and $\underline j$ is the
derivative part of the charged current
\begin{eqnarray}
{\underline J} ={\underline j}
-\frac{e}{mc}\rho {\underline A}
\label{curr} \end{eqnarray}
that enters in the equation for charge
conservation.\parno
To derive the Kubo formula for conductivity \cite{Mah1},
one perturbes the Hamiltonian
with a term $ \delta H = {1\over c}\int d^3x \delta {\underline A}^{ext}
(\underline x,t)\cdot {\underline J}(\underline x)$,
which couples  the total  current to a weak external electric field.
Linear response gives the conductivity tensor. Under the assumption
that the Hamiltonian (\ref{(0.1)}) is {time-independent}
the response is
function of the time difference only. Its Fourier transform is
\begin{eqnarray}
\sigma_{ij} (\underline x, \underline x^\prime, \omega ) =
{i\over {\hbar\omega }}
\pi^{\rm Ret}_{ij} (\underline x,\underline x^\prime,
\omega)- {{  i\, e}\over {m\omega}}
\langle \rho (\underline x) \rangle \delta_{ij} \delta_3
(\underline x-\underline x^\prime)
\label{(2)}
\end{eqnarray}
where
$i,j$ are space directions, $\pi^{\rm Ret}_{ij}(\omega)$
is the connected retarded current-current correlator.
This is going to be the object of our discussion.
\parno
It is
convenient to use imaginary time ($it=\tau$) and to do perturbation
theory in terms of time-ordered Green functions
\begin{eqnarray}
\pi^{\rm C}_{\mu\nu }(\underline x,  \underline x^\prime,\tau- \tau^\prime )=
- \langle  {\cal T} J_\mu(\underline x,\tau)J_\nu (\underline x',
\tau^\prime )\rangle
\label{(3)}
\end{eqnarray}
where the four-dimensional notation means $J_\mu= (c\rho,
\underline J), \,\, \mu=0,1,2,3$ (
since we use the Matsubara formalism, the metric is Euclidean).
Brackets indicate equilibrium
thermal average. The relation between the two
kinds of correlators is a standard result \cite{Ric} and it is
briefly recalled here. The Fourier transform of $\pi^{\rm C}_{\mu\nu }
(\underline x,  \underline x^\prime,\tau)$ is a sum over a discrete
number of frequencies (Matsubara) due to the periodicity
under the shift $\tau\to \tau + \hbar\beta$
\begin{eqnarray}
\pi^{\rm C}_{\mu\nu }(\underline x,  \underline x^\prime,\tau)
=
{1\over {\hbar\beta}}\sum_n
e^{-i\omega_n\tau} \pi^{\rm C}_{\mu\nu}
({\underline x},{\underline x}^\prime, \omega_n).
\label{(3.1)}
\end{eqnarray}
The correlator of interest $\pi^{\rm Ret}( \omega)$ can be written in terms
of $\pi^{\rm C}( \omega_n)$ by analytic continuation. For real $\omega$:
\begin{eqnarray} &&
\pi^{\rm Ret}_{\mu\nu}( \omega)
= \int^\infty_{-\infty} d\omega'
\frac{f_{\mu\nu}(\omega')}{\omega-\omega'+i\epsilon}
\nonumber\\&&
\pi^{\rm C}_{\mu\nu}( \omega_n)
= \int^\infty_{-\infty} d\omega' \frac{f_{\mu\nu}(\omega')}{i\omega_n-\omega'}
\label{(3.2)}
\end{eqnarray}

A delicate issue is the presence of disorder. We can consider
various approximations depending on the model describing the
experimental device. One possibility is that the disorder
is described by impurities and that the averaging is performed
on the final result, thus allowing a correlation between the
layers. This approach can be simplified by the independent
averaging on the two layers.
%A stronger assumption consists in assuming that the temperature
%is sufficient to take into account the {dissipation}
%present in the systems.
In the present approach we
follow the last point of view in order to see the consequences
of the direct elecromagnetic interaction of the two layers.
This simplifying dynamics has the consequence that translational
invariance is valid in the plane ${\bf r}=(x,y)$. Confinement
is in the $z$ direction. We then Fourier transform with a bidimensional
wave-vector ${\bf q}$. \parno
The static and homogeneous limit of the  transconductivity
is:
\begin{eqnarray} \sigma_{ij} (z,z^\prime) =
\lim_{\omega \to 0} \lim_{q\to 0}
%\left (
{i\over {\hbar\omega}}\,\pi^{\rm Ret}_{ij}
({\bf q}, z, z^\prime,\omega )
%\right)
\label{(4)}
\end{eqnarray}
where $z$ and $z^\prime $ belong to different layers.
{Moreover} in the limit of thin layer we have to impose
\begin{eqnarray}
\sigma_{3 j}  = \sigma_{j 3} = 0,
\label{(4.1)}
\end{eqnarray}
because there is no flow of charge in the z-direction. Also
the conservation of the electromagnetic current
requires the above condition in the limit of thin {layers.}
\parno
In the next section, the current-current correlator, and therefore
the conductivity,
will be related to the polarization tensor of the e.m. field.
\vskip 0.5truecm

\section{The polarization tensor}
\label{sec:polar}
The thermal Green functions for the photon field  are
\begin{eqnarray} {\cal D}_{ij}(\underline x,\tau, \underline x^\prime,\tau^\prime)=
-{1\over{\hbar}}\langle {\cal T}A_i(\underline x, \tau)
A_j(\underline x^\prime,\tau^\prime)\rangle \label{(5)} \end{eqnarray}
The prefactor $1/\hbar $ ensures  the same
dimension as the Coulomb interaction ${\cal D}_{00}$.  The
propagators for free photons and the bare Coulomb interaction are
best written in momentum space:
\begin{eqnarray} {\cal D}^{(0)}_{ij}(\underline k,\omega_n)= - \left (
\delta_{ij}- { {k_i k_j}\over {k^2} }  \right ){{4\pi c^2}\over
{\omega_n^2+c^2k^2}} ,\quad {\cal D}^{(0)}_{00}({\underline k},\omega_n )=
{{4\pi}\over {k^2}} \label{(6)} \end{eqnarray}
and are the components of a tensor ${\cal D}^{(0)}_{\mu\nu}$,
with $ {\cal D}^{(0)}_{0i}={\cal D}^{(0)}_{i0}=0$. When the interaction with matter
is included, the dressed photon propagator and the effective Coulomb interaction
are components of a tensor ${\cal D}_{\mu\nu}$ which differs from the bare
one by polarization insertions:
\begin{eqnarray}
{\cal D}_{\mu\nu}(\underline x,\underline x^\prime,\omega_n)
&-& {\cal D}^{(0)}_{\mu\nu}(\underline x,\underline x^\prime,\omega_n)
\nonumber\\&
=&{\cal D}^{(0)}_{\mu\rho}(\underline x,\underline x_1,\omega_n){\cal P}_{\rho
\sigma}(\underline x_1,\underline x_2,\omega_n){\cal D}^{(0)}_{\sigma\nu}(
\underline x_2, \underline x^\prime ,\omega_n )
\label{(7)}
\end{eqnarray}
Summation and integration of repeated variables are understood hereafter. The
identification of the polarization insertion of the Coulomb interaction
with the connected density-density correlator is well known
from textbooks:
\begin{eqnarray}
{\cal P}_{00}(\underline x,\underline x^\prime ,\omega_n)=
{1\over{\hbar c^2}}\pi^{\rm C}_{00}
(\underline x,\underline x^\prime,\omega_n)
\label{(8a)}
\end{eqnarray}
A perturbative analysis shows the further exact relations among the
polarization insertion of the photon propagator and the current-current
correlator:
\begin{eqnarray} {\cal P}_{ij}(\underline x, \underline x^\prime , \omega_n)=
-{e\over mc^2 }\langle \rho(\underline x)\rangle
\delta_{ij}\delta_3(\underline x-\underline x^\prime) +
{1\over {\hbar c^2}}\pi^{\rm C}_{ij}(\underline x, \underline x^\prime,\omega_n)
\label{(8b)}
\end{eqnarray}
\begin{eqnarray} {\cal P}_{0i}={1\over {\hbar c^2}}\pi^{\rm C}_{0i},\quad {\cal P}_{i0}=
{1\over {\hbar c^2}} \pi^{\rm C}_{i0} \label{(8c)} \end{eqnarray}
Therefore, the  conductivity tensor is  proportional to the retarded
photon polarization:
\begin{eqnarray}  \sigma_{ij} (\underline x, \underline x^\prime, \omega ) =
i{{c^2}\over \omega} {\cal P}^{\rm Ret}_{ij}(\underline x, \underline x^\prime,
\omega)
\label{(9)}
\end{eqnarray}
In our approximation, after the average on disorder, both $\sigma $
and $\cal P$ are functions of the difference ${\bf r}-{\bf r^\prime }$,
which is traded for the variable ${\bf q}$. The dependence in $z$ and
$z^\prime $ will be simplified with the limit of thin layers.
\vskip 0.5truecm

\section{The Dyson equation}
\label{sec:dyson}
To evaluate in some approximation the polarization, we start by
writing  an exact Dyson equation in coordinate space. Due to the
geometry of the problem, we find it convenient to rearrange the graphs
according to the two layer configuration:
\begin{eqnarray}
{\cal P}_{\mu\nu}(\underline x, \underline x^\prime, \omega_n)&&=
{\cal P}^\star_{\mu\nu}(\underline x, \underline x^\prime, \omega_n)
\nonumber\\&&
+
{\cal P}^\star_{\mu\rho}(\underline x, \underline x^{\prime\prime},\omega_n)
{\cal D}^{(0)}_{\rho\sigma}(\underline x^{\prime\prime} -
\underline x^{\prime\prime\prime}, \omega_n)
{\cal P}_{\sigma\nu}(\underline x^{\prime\prime\prime},
\underline x^\prime, \omega_n)
\label{(10)}
\end{eqnarray}
where ${\cal D}^{(0)}_{\rho\sigma}$ is an interlayer bare propagator
(points in different layers) and ${\cal P}^\star $ is the
irreducible polarization tensor, given by the sum of e.m. polarization
insertions that cannot be disconnected by
cutting a single interlayer photon or Coulomb line.\parno
Next we take the average over disorder, and make the simplification of
retaining the same Dyson equation for averaged correlators. In such
approximation, the equation is written in ${\bf q}$-space:
\begin{eqnarray}
{\cal P}_{\mu\nu}({\bf q}, z, z^\prime , \omega_n) =&&
{\cal P}^\star_{\mu\nu}({\bf q}, z, z^\prime , \omega_n)
\\
+&& {\cal P}^\star_{\mu\rho}({\bf q}, z,z^{\prime\prime}, \omega_n)
{\cal D}^{(0)}_{\rho\sigma}({\bf q},
z^{\prime\prime}-z^{\prime\prime\prime}, \omega_n)
{\cal P}_{\sigma\nu}({\bf q},z^{\prime\prime\prime},z^\prime, \omega_n)
\nonumber
\label{(11)}
\end{eqnarray}
We put $\underline k = ({\bf q}, k_3)$ in the bare e.m. propagators
and antitransform to $z-z^\prime$:
\begin{eqnarray}
{\cal D}^{(0)}_{\mu\nu}({\bf q}, z-z^\prime ,\omega_n) =
\int_{-\infty}^\infty {{dk_3}\over {2\pi}}
{\cal D}^{(0)}_{\mu\nu}(\underline k,\omega_n)
e^{ik_3(z-z^\prime)}
\label{(12)}
\end{eqnarray}
With the definition of the auxiliary function
\begin{eqnarray} {\tilde{\cal D}}({\bf q},z-z^\prime,\omega_n ) =
 2 \pi c {{e^{-{1\over c}|z-z^\prime |\sqrt {\omega_n^2 +
q^2c^2} } } \over {\sqrt {\omega_n^2 + q^2c^2} } }\label{(13)} \end{eqnarray}
we evaluate
\begin{eqnarray} &&{\cal D}^{(0)}_{00} =
2\pi {{e^{-q|z-z^\prime|}}\over q}
\nonumber\\&&
{\cal D}^{(0)}_{ab}= - \delta_{ab} {\tilde {\cal D}} - {{q_aq_bc^2}\over
{\omega_n^2}}( {\tilde{\cal D}}- {\cal D}^{(0)}_{00})
\nonumber\\&&
D^{(0)}_{33} = {{ q^2c^2}\over {\omega_n^2}} ({\tilde {\cal D}}-
{\cal D}^{(0)}_{00})
\nonumber\\&&
{\cal D}^{(0)}_{3a}= {\cal D}^{(0)}_{a3} = 2\pi i \,{\rm {sign}}(z-z^\prime )
{{q_ac^2}\over {\omega_n^2}}
\left ( e^{-{1\over c}|z-z^\prime|\sqrt {\omega_n^2 + c^2q^2} } -
e^{-q|z-z^\prime |} \right )
\nonumber\\&&
{\cal D}^{(0)}_{0j}= 0,
\label{(15)}
\end{eqnarray}
where the last equation is imposed by the gauge fixing condition
$\partial_jA_j=0$.
To proceed further, we consider the limit of thin layers.
Since the confining well in the z-direction causes a strong splitting
in the energy eigenvalues and the dynamics is translation
invariant in the plane $z = {\rm const}$,
we allow only single particle states of the form
(for the unperturbed basis of the Hilbert space)
\begin{eqnarray}
u_\ell({\bf r},z) = \phi_\ell(z)\psi_\ell({\bf r})
\label{(15.1)}
\end{eqnarray}
The motion in
z-direction is thus frozen and all bilinear local operators contain a
factor $|\phi_\ell(z)|^2$. The limit of thin layer is performed
by
\begin{eqnarray}
 |\phi_\ell(z)|^2 \to \delta(z-z_\ell).
\label{(15.2)}
\end{eqnarray}
It is customary
to introduce two canonical basis of fermion operators that create or
remove a particle with position $(x,y)={\bf r}$ and spin $\sigma $ in
layer $\ell =1,2$. With $z_1=0$ and $z_2=d$, we have the relevant
limit operators:
\begin{eqnarray}
\rho (\underline x)=  \sum_\ell \rho_\ell ({\bf r})\delta(z-z_\ell),
\quad \rho_\ell ({\bf r})= -e\sum_\sigma \psi^\dagger_{\ell\sigma}({\bf r} )
\psi_{\ell\sigma}({\bf r}) \label{(16a)}
\end{eqnarray}
\begin{eqnarray}
J_a (\underline x) = \sum_\ell J_{a\ell} ({\bf r)} \delta (z-z_\ell), \quad
J_{a\ell} ({\bf r})= j_{a\ell} ({\bf r})- {e\over {mc}}\rho_\ell
({\bf r}) A_a ({\bf r}, z_\ell) \label{(16b)}
\end{eqnarray}
\begin{eqnarray}
j_{a\ell}({\bf r}) =i{{\hbar e}\over {2m}}
\sum_\sigma \psi^\dagger_{\ell\sigma}
({\bf r})\partial_a \psi_{\ell\sigma}({\bf r}) -
(\partial_a \psi^\dagger_{\ell\sigma}({\bf r}))\psi_{\ell\sigma} ({\bf r})
\end{eqnarray}
In the thin layer limit the component $J_3$ is assumed to vanish.
\par\noindent
The index $a$ will hereafter denote transverse space components $(x,y)$,
while $i$ is used for components $(x,y,z)$, and  a greek index
includes the imaginary time component.\parno
The absence of tunnelling ensures the  conservation of charge
in each layer, which reads:
\begin{eqnarray} {1\over {i\hbar}}[H,\rho_\ell ({\bf r})] = {\rm div}_{xy}\,
J_\ell({\bf r} ).
\label{(17)}
\end{eqnarray}
In the thin layer limit the Dyson equations (11) become algebraic, with all
delta functions factoring out or allowing to make the double integrations
in the $z$ direction:
\begin{eqnarray} {\cal P}_{\mu\nu}(\ell\ell^\prime)-
{\cal P}^\star_{\mu\nu}(\ell\ell^\prime)=
{\cal P}^\star_{\mu\rho}(\ell\ell_1){\cal D}^{(0)}_{\rho\sigma}(\ell_1\ell_2)
{\cal P}_{\sigma\nu}(\ell_2\ell^\prime ) \label{(18)} \end{eqnarray}
The variables ${\bf q}$ and $\omega_n$ are omitted for brevity. Recall
that the Dyson equation was
constructed with the requirement that ${\cal D}^{(0)}$ connects different
layers, then $\ell_1\neq \ell_2$. In the formula we put
\begin{eqnarray} {\cal D}^{(0)}_{\mu\nu} (\ell\ell^\prime)= {\cal D}^{(0)}_{\mu\nu}({\bf q},
z_\ell- z_{\ell^\prime},\omega_n), \quad
{\cal P}_{\mu\nu}(\ell,\ell^\prime )={\cal P}_{\mu\nu}({\bf q},
\omega_n, z_\ell,z_{\ell^\prime} ) \label{(19)} \end{eqnarray}
They are respectively the entries of two  $4\times 4$ matrices
${\bf D}^{(0)}(\ell\ell^\prime )$ and ${\bf P}(\ell\ell^\prime)$. Therefore,
the  Dyson equations correspond to 4 matrix equations $(\ell,\ell^\prime=1,2)$:
\begin{eqnarray}{\bf P}(\ell\ell^\prime)={\bf P}^\star (\ell\ell^\prime) +
{\bf P}^\star (\ell 1){\bf D}^{(0)}(1 2){\bf P}( 2 \ell^\prime )+
{\bf P}^\star (\ell 2){\bf D}^{(0)}(2 1){\bf P}( 1 \ell^\prime )
\label{(20)} \end{eqnarray}
The structure of the polarization tensor is greatly limited by symmetry
and charge conservation. The latter implies the following exact relations:
\begin{eqnarray}
i{{\omega_n}\over c}{\cal P}_{0\nu} ({\bf q},\omega_n,\ell,\ell^\prime)=
q_a {\cal P}_{a\nu} ({\bf q},\omega_n,\ell,\ell^\prime), \quad \nu=0,1,2,3
\label{(21)}
\end{eqnarray}
The same holds when the indices are exchanged. These relations
correspond to the Ward identity relating the vertex functions for
Coulomb and e.m. coupling to the electron field. In absence of
external magnetic field rotational symmetry requires
the tensor structure
\begin{eqnarray}
{\cal P}_{ab} ({\bf q}, \omega_n, \ell,\ell^\prime) =
\delta_{ab}A({\bf q},\omega_n, \ell,\ell^\prime) + {{q_aq_b}\over {q^2}}
B ({\bf q},\omega_n, \ell,\ell^\prime) \label{(22)}
\end{eqnarray}
From charge conservation we find:
\begin{eqnarray} {\cal P}_{0a}={\cal P}_{a0}=
i{{\omega_n}\over {cq^2}}q_a {\cal P}_{00},
\quad
A+B=-{{\omega_n^2}\over {c^2q^2}}{\cal P}_{00}. \label{(23)}
\end{eqnarray}
\section{The Interlayer Polarization in RPA}
\label{sec:inter}
We shall solve Dyson's equation in RPA and for identical layers.
In this approximation ${\bf P}^\star$ is deprived of all interlayer
interaction lines and therefore
\begin{eqnarray} {\bf P}^\star (\ell\ell^\prime)=
\delta_{\ell\ell^\prime}
{\bf P}^{(0)}_\ell
\label{(24)} \end{eqnarray}
where {${\bf P}^{(0)}_\ell $} is the exact polarization matrix of the
single isolated layer with its internal dynamics. The index $\ell$ keeps
{track} of the charge of the carriers. However in the present
approximation scheme (no charged impurities and no interlayer
interaction) the charge appears always with even powers. Thus the index
$\ell$ can be neglected.\parno
With this simplification, Dyson's equations eqs (\ref{(20)}) are:
\begin{eqnarray}
{\bf P}(12) =&& {\bf P}^{(0)}{\bf D}^{(0)}(12){\bf P}(22)
\nonumber\\
{\bf P}(22) =&& {\bf P}^{(0)}+{\bf P}^{(0)}{\bf D}^{(0)}(21){\bf P}(12).
\label{(25)}
\end{eqnarray}
After the analytic continuation, the  sub-matrix  $P(22)_{ab}$
is proportional to the conductivity tensor of layer 2,
while $P(12)_{ab}$ is proportional to transconductivity among layers 1
and 2.
Eqs. (\ref{(25)}) states that these two quantities are related
by the electromagnetic interaction.\par
\par

The various components of ${\bf P}^{(0)}$ fulfill the relations
of rotational
symmetry and charge conservation, such as (index $\ell$ is here forgotten)
\begin{eqnarray} &&
P^{(0)}_{ab}= A^{(0)}\delta_{ab}+B^{(0)} {{q_aq_b}\over q^2}, \quad
P^{(0)}_{0a}=P^{(0)}_{a0}= iq_a{{\omega_n}\over {q^2c}}P^{(0)}_{00},
\nonumber\\ &&
A^{(0)}+B^{(0)}={{q_aq_b}\over {q^2}} P^{(0)}_{ab} =
-{{\omega_n^2}\over {c^2q^2}}
P^{(0)}_{00}.
%\nonumber\\
%&&P^{(0)}_{a3}= P^{(0)}_{3a}
%= i{{\omega_n}\over {cq^2}}q_a  P^{(0)}_{03},
%\quad
% P^{(0)}_{03}=P^{(0)}_{30}.
\label{(26)}
\end{eqnarray}
$P^{(0)}_{ab}$ is directly linked to the conductivity tensor
$\sigma^{(0)}_{ab}$ of the isolated layer. \parno
The coupled Dyson  equations (25) provide a matrix equation for the
interlayer polarization:
\begin{eqnarray} {\bf P}(12)= {\bf Q} + {\bf Q}{\bf D}^{(0)}(21){\bf
P}(12)
\label{(27)}
\end{eqnarray}
where
${\bf Q}={\bf P}^{(0)}{\bf D}^{(0)} (12){\bf P}^{(0)}$.\parno
Before solving the equation, let us evaluate the components $Q_{ab}$,
which give the lowest order approximation in the interlayer interaction
to transconductivity. From eq. (\ref{(15)}) we get
\begin{eqnarray}
Q_{ab}={ P}^{(0)}_{a0}{\cal D}^{(0)}_{00}(12){ P}^{(0)}_{0b}+
{ P}^{(0)}_{ac}{\cal D}^{(0)}_{cd}(12){ P}^{(0)}_{db}.
\label{(27.1)}
\end{eqnarray}
By using charge conservation for the polarization we obtain after
some algebra
\begin{eqnarray}
Q_{ab}=&&-{\tilde{\cal D}}\left[A^{(0)2} (\delta_{ab}
-{{q_a q_b}\over {q^2}})+
{{q_aq_b}\over {q^2}}(P^{(0)}_{00})^2
{{\omega_n^2}\over {c^2q^4}}(\omega_n^2+c^2q^2)\right]
%\nonumber\\&&
%+\left({{i\omega_n}\over {cq^2}}\right)^2
%q_aq_b P^{(0)2}_{03} {\cal D}^{(0)}_{33}
\label {(28)}
\end{eqnarray}
where ${\tilde{\cal D}}$ is the function in eq. (\ref{(13)})
with $|z-z^\prime |=d$.
From the isotropic part of $Q_{ab}$ we obtain transresistance to
first order in interlayer interaction: $\sigma(12) =
(2\pi/c) \sigma^{(0)2} $.
The experiments are carried under the condition that no current
flows in the driven layer. The measured quantities are the driving current
$J^{(1)}$ and the electric field $E^{(2)}$ that builds up in layer two to
balance the drag field. Transresistance is the ratio
\begin{eqnarray}
\rho (12) = {{E^{(2)}}\over {J^{(1)} }} = -{{\sigma (12)}\over {\sigma (11)
\sigma (22)-\sigma (12)\sigma (21) }}
\end{eqnarray}
Under the condition $\sigma (12)<<\sigma^{(0)}$, we approximate the layer conductance
$\sigma (\ell\ell)$
in presence of the other by the value $\sigma^{(0)}$ of the isolated layer.

We then find  $\rho (12)\approx
\sigma(12)/\sigma^{(0)2}$ and obtain a universal result for
transresistivity:
\begin{eqnarray}
\rho (12)= {{2\pi}\over c} = \alpha R_H
\label {(29)}
\end{eqnarray}
where $R_H={\rm h}/e^2$ is the Hall resistance and $\alpha $ is the fine
structure constant corrected for the interlayer medium.\par
While ${\bf Q}$ represents the inter-layer polarization with a single
one-photon exchange, the solution of \ref{(27)} is the polarization with
an effective interlayer interaction, in RPA. ${\bf P}^{(0)}$ corresponds
to disorder averaged single layer polarization where the second layer is
absent.
If we consider also the second layer, then our approximation neglects
those contributions to the disorder average that correlate various
${\bf P}^{(0)}$ insertions. This last approximation
was used also in the evaluation of Coulomb and phonon drag.
One computes:
\begin{eqnarray}
P(12)_{00}({\bf q}, \omega_n)=
{{ {P^{(0)}_{00}}^2({\bf q},\omega_n) {{2\pi}\over {cq^2}}\sqrt
{\omega_n^2+c^2q^2}
e^{-{d\over c}\sqrt {\omega_n^2+c^2q^2}}}\over
{1-  {P^{(0)}_{00}}^2 {{4\pi^2}\over {c^2q^4}}(\omega_n^2+c^2q^2)
e^{-2{d\over c}\sqrt {\omega_n^2+c^2q^2}}
%+ {{4\pi^2c^2}\over {\omega_n^2}}
%P^{(0)}_{00} P^{(0)}_{33} (e^{-{d\over c}\sqrt {\omega_n^2+c^2q^2}}
%- e^{-qd})^2
}}
\label {(30)}
\end{eqnarray}
In the diffusive regime, we use
\begin{eqnarray}
P^{(0)}_{00}({\bf q}, \omega) = \Gamma e^2 {{Dq^2}\over {Dq^2 - i\omega }}
\label {(31)}
\end{eqnarray}
where $D$ is the diffusion constant and $\Gamma $ is related to the single
layer conductivity: $\sigma^{(0)}=\Gamma e^2 D$.
%$P_{33}$ is a constant.
In this approximation, the limits $q\to 0$, $\omega\to 0$  of
trans-conductivity yield a small correction to (29) for transresistance:
\begin{eqnarray} \rho (12)  = {{2\pi}\over c} {1\over
{1-\sigma^{(0)2} (2\pi/c)^2}}.
\label {(32)} \end{eqnarray}
%%%%%%%%%%%%%%%%%%%%%%%%%%%%%%%%%%%%%%%%%%%%%%%%555555
%%%%%%%%%%%%%%%%%%%%%%%%%%%%%%%%%%%%%%%%%%%%%%%%555555
%%%%%%%%%%%%%%%%%%%%%%%%%%%%%%%%%%%%%%%%%%%%%%%%555555
\section{Discussion}
In the present work we have considered the virtual photon
contribution to the frictional drag in two parallel
layers. In contrast to the Coulomb drag the contribution
of the single virtual photon survives in the limit
of $q \to 0$ and $\omega \to 0$. The RPA approximation allows
to get the result in terms of the full single layer
polarization even in presence of disorder.
The result is valid also at finite temperature, which enters
in the single layer conductivity.
Eq. (\ref{(28)}) gives the contribution of
one single-photon-exchange and eq. (\ref{(30)}) yields the
sum over all possible single-photon-exchanges. The result
is large, and seems not to compare with the experimental values.
This is peculiar to the e.m. contribution: the  idealization of thin
and infinite layers is the same used in previous works devoted to
Coulomb or phonon drag. We also made the same approximation of neglecting
correlations due to disorder among polarization insertions in the
evaluation of the effective interaction.\parno
This peculiarity goes along with that of being unscreened, and the possibility
of significative  changes in more realistic models, due to finite size effects.
We leave this analysis to a future investigation.
\par\noindent

\section*{Acknowledgements}
Two of us (R.F. and L.M.) wish to thank Vincenzo Benza for stimulating
discussions.

\begin{center}

\begin{figure}

\begin{center}

\includegraphics[width=10cm,keepaspectratio]{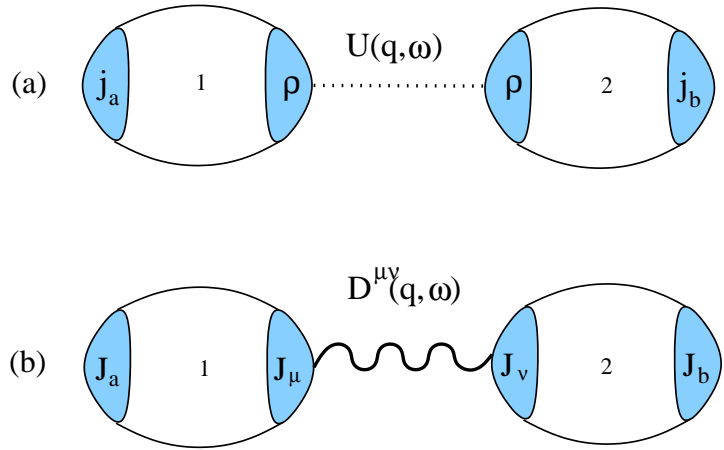}

\caption{The first order contribution to the current-current
correlator: coupling mediated by (a) screened Coulomb interaction,
(b) four-dimensional electromagnetic propagator. $\rho$ is the
electron density, ${\bf j}$ the paramagnetic current density and
${\bf J} = {\bf j} - {{(-e)} \over {mc}}\rho {\bf A}$. The number
inside the bubbles indicates the subsystem. The intralayer
interaction is included within RPA in each bubble. (a) vanishes in
the DC limit and corresponds to the $\mu=\nu=0$ component of (b).
The space-like components of (b) are \emph{nonvanishing} in this
limit.}

\end{center}

\end{figure}

\end{center}

%%%%%%%%%%%%%%%%%%%%%%%%%%%%%
\vfill
\end{document}